\documentclass[11pt]{article}
\usepackage[margin=1in]{geometry}
\usepackage{graphicx}
\usepackage{hyperref}
\usepackage[numbers,sort&compress]{natbib}
\usepackage{amsmath}
\usepackage{amssymb}
\usepackage{enumitem}
\usepackage{booktabs}
\usepackage{xcolor}
\hypersetup{colorlinks=true,linkcolor=blue,citecolor=blue,urlcolor=blue}

\newcommand{\pkg}[1]{\texttt{#1}}

\title{Coord2Region: A Python Package for Mapping 3D Brain Coordinates to Atlas Labels, Literature, and AI Summaries}
\author{
Hamza Abdelhedi$^{1,2,\star}$\thanks{Corresponding author: \texttt{hamza.abdelhedii@gmail.com}},
Yorguin-Jose Mantilla-Ramos$^{1,2,\star}$,
Sina Esmaeili$^{1,\star}$, \\
Annalisa Pascarella$^{3}$,
Vanessa Hadid$^{4}$,
Karim Jerbi$^{1,2,5}$\\[0.5ex]
\small $\star$ These authors contributed equally.\\[1ex]
\small $^{1}$ Cognitive and Computational Neuroscience Laboratory (CoCo Lab), Psychology Department, \\
University of Montreal, Montreal, QC, Canada\\
\small $^{2}$ Mila – Quebec Artificial Intelligence Institute, Montreal, QC, Canada\\
\small $^{3}$ Institute for Applied Mathematics Mauro Picone, National Research Council, 00185 Rome, Italy\\
\small $^{4}$ Psychology Department, University of Montreal, Montreal, QC, Canada\\
\small $^{5}$ UNIQUE Center (Quebec Neuro-AI Research Center), Montreal, QC, Canada
}\date{\today}
\begin{document}
\maketitle

\begin{abstract}
\noindent
We present Coord2Region, an open-source Python package that streamlines coordinate-based neuroimaging workflows by automatically mapping 3D brain coordinates (e.g., MNI \cite{evans19933d} or Talairach \cite{talairach1980application}) to anatomical regions across multiple atlases. The package links mapped coordinates to meta-analytic resources via the Neuroimaging Meta-Analysis Research Environment (NiMARE) \cite{salo2022nimare}, providing direct integration with Neurosynth \cite{yarkoni2011large} and NeuroQuery \cite{dockes2020neuroquery}. This directly connects coordinates and regions to the broader neuroimaging literature. In addition to atlas-based labeling and literature retrieval, Coord2Region offers an optional large language model (LLM) functionality that generates text summaries of linked studies and illustrative images of queried regions. These AI-assisted features are intended to support interpretation and exploration, while remaining clearly complementary to peer-reviewed literature and established neuroimaging tools. Coord2Region provides a unified pipeline with a robust command-line interface, flexible dataset management, and provider-agnostic LLM utilities, and it supports both single-coordinate and high-throughput batch queries with nearest-region fallback for volume and surface atlases. Furthermore, Coord2Region includes a web interface for interactive configuration (via JSON Schema forms) and cloud execution (via Hugging Face), enabling users to build YAML configurations and run analyses in-browser without local installation. Together, these capabilities lower friction, reduce manual errors, and improve reproducibility in coordinate-centric neuroimaging workflows, promoting more robust and transparent research practices.

\end{abstract}

\section{Statement of Need}
Neuroimaging studies frequently report activation locations using standardized 3D coordinates (e.g., MNI \citep{evans19933d} or Talairach \citep{talairach1980application}). However, translating these coordinates into consistent anatomical and functional labels remains error-prone and time-consuming. GUI viewers such as MRIcroGL \citep{rorden2025mricrogl, MRIcroGL_NITRC} and MRIcron \citep{rorden2000stereotaxic, MRIcron_NITRC} are excellent for interactive image exploration and rendering, but they do not support automated, programmatic coordinate-to-label mapping. Web-based viewers like NiiVue \citep{NiiVue_Docs, Drake2021NiiVue} broaden access but likewise prioritize visualization over batch labeling. Existing tools such as Talairach Software \citep{Lancaster2000TalairachDaemon} and label4MRI \citep{Chuang2022label4MRI} are outdated, platform-dependent, or limited in their atlas coverage, leaving researchers to manually translate coordinates into brain regions—a practice that introduces inconsistency and undermines reproducibility. Meta-analytic toolchains such as Neurosynth and NeuroQuery bridge text and brain maps \citep{yarkoni2011large, dockes2020neuroquery}, and NiMARE unifies modern meta-analytic methods \citep{salo2022nimare}, yet researchers still spend substantial time searching the literature to interpret the functional significance of reported regions. In short, there is a gap between isolated coordinates and consistent, scriptable labels enriched with literature context.

\noindent
Coord2Region was created to address this gap by providing a user-friendly, automated Python tool for coordinate-to-region mapping across multiple brain atlases that also links atlas lookups to meta-analytic resources for supporting literature. The package offers a unified pipeline and command-line interface (CLI), multi-atlas handling, KD tree–accelerated searches, large language model (LLM)–based region summaries and images, and reproducible exports. In addition, Coord2Region provides a web interface with a JSON Schema–based configuration builder that generates valid YAML files and a cloud runner (via Streamlit) that can execute the pipeline directly in the browser, making the system usable even without local installation. Built on established open-source neuroimaging libraries (nibabel \citep{nibabel}, nilearn \citep{Nilearn, abraham2014machine}, and MNE-Python \citep{MNE_Python, Gramfort2013}) and supporting multiple LLM APIs (Gemini, OpenAI, Hugging Face), Coord2Region integrates readily into standard research workflows, reducing manual effort and errors while enabling reproducible neuroscience.

\paragraph{Contributions and strengths.} 
\begin{itemize}
    \item \textbf{Automated anatomical labeling.} Automated anatomical region identification across more than 20 atlases, eliminating manual lookups and improving consistency across studies. \emph{For example, a researcher with 150 fMRI activation peaks can run a single batch query and obtain standardized region names from multiple atlases in seconds.}

    \item \textbf{Batch processing at scale.} Efficient batch handling of large datasets of 3D coordinates. For instance, lesion coordinates from hundreds of patients can be mapped automatically without interactive software.

    \item \textbf{Cross-atlas robustness.} Atlas-agnostic, KD-tree–accelerated nearest-region fallback (for both volume and surface atlases), ensuring that every coordinate receives a label while reporting distances to support interpretation of uncertainty.

    \item \textbf{Literature linkage.} Integration with Neurosynth and NeuroQuery via NiMARE \citep{yarkoni2011large, dockes2020neuroquery, salo2022nimare}, linking atlas-defined regions directly to more than 20{,}000 published functional neuroimaging studies. \emph{For example, a coordinate in the left inferior frontal gyrus can be associated with studies reporting its role in language processing.}

    \item \textbf{Optional LLM utilities.} Provider-agnostic LLM utilities (synchronous, asynchronous, and streaming) for generating text summaries of linked studies and illustrative images of queried regions. These outputs are explicitly positioned as aids to interpretation, not replacements for systematic review.

    \item \textbf{Unified access modes.} A single \pkg{run\_pipeline} API and CLI that implement the full coordinates$\rightarrow$atlas$\rightarrow$literature$\rightarrow$AI-summary workflow. All functionality is exposed through the Python API, a command-line interface, and a reproducible YAML configuration system, enabling use by both developers and non-programmers.

    \item \textbf{Web interface.} A web interface for building configuration files, generating CLI commands from user-specified inputs, outputs, and options, and a cloud runner (via Streamlit / Hugging Face) that executes the end-to-end workflow in the browser.

    \item \textbf{Flexible applications.} Flexible usage across modalities and paradigms, including fMRI, EEG/MEG source localization, lesion mapping, and intracranial electrode studies.
    
    \item \textbf{Open science practices.} Open-source and actively maintained on GitHub, with a community-driven development model that welcomes contributions, alongside rich documentation, examples, and unit tests to support reproducible use and extension.
\end{itemize}

\section{Methods}
\subsection{Overview}

Coord2Region adopts a modular architecture designed for flexibility, maintainability, and ease of extension. Each module fulfills a distinct role, yet the components can be composed into a unified pipeline. Given (a) one or more coordinates or (b) a region name, the pipeline executes the following steps:

\begin{enumerate}[label=\textbf{M\arabic*}.]
    
    \item \textbf{Data initialization.} Resolve directories, load available atlases, precompute centroids, build KD-trees, and load metadata. This initialization ensures that subsequent mapping steps are efficient and reproducible. \emph{Example: when running multiple queries in a session, atlas files are loaded once and cached for reuse, reducing runtime from minutes to seconds.}

    \item \textbf{Coordinate-to-region mapping.} Translate input coordinates into anatomical labels using \pkg{AtlasMapper} or \pkg{MultiAtlasMapper}. If a coordinate falls outside labeled regions, a nearest-region search is performed using prebuilt KD-trees, and the Euclidean distance to the nearest voxel or vertex is reported. \emph{Example: a peak activation at $(-42, -22, 8)$ can be mapped to ``left inferior frontal gyrus,'' with a 2.3~mm distance to the nearest atlas voxel.}

    \item \textbf{Study retrieval.} Retrieve studies around a coordinate $x$ within radius $r$ or by region label using NiMARE-backed datasets (Neurosynth / NeuroQuery) with CrossRef / PubMed metadata fill. \emph{Example: the same inferior frontal gyrus coordinate can return dozens of linked studies, including work on language processing and cognitive control.}

    \item \textbf{Summarization and visualization (optional).} Use LLMs via \pkg{AIModelInterface} (a provider-agnostic wrapper) for generating text summaries of linked studies and optional AI-based images of queried regions. Coord2Region can also produce atlas overlays for MNI coordinates with watermarking. These features help researchers rapidly contextualize results, while remaining clearly complementary to systematic literature review.

    \item \textbf{Export.} Save results in multiple formats (JSON, CSV, PDF, pickle, or directory trees), ensuring compatibility with downstream analyses and facilitating reproducibility.

\end{enumerate}

\subsection{Atlas mapping}
This component forms the core of Coord2Region. 

\noindent
\pkg{AtlasFetcher \& AtlasFileHandler.} classes handle retrieval and I/O for brain atlas datasets, supporting downloads from online repositories and loading from local files (e.g., NIfTI \texttt{.nii/.nii.gz}, NumPy \texttt{.npz}). Together, they streamline atlas management and enable rapid integration of new atlases, improving scalability and usability. \emph{Practical use:} a researcher can download and register multiple atlases (e.g., AAL, Harvard--Oxford, Destrieux) in a single call and apply them across the same dataset.

\noindent
\pkg{AtlasMapper, MultiAtlasMapper, BatchAtlasMapper.} These core classes perform coordinate-to-region mapping, with complementary scopes:
\begin{itemize}
    \item \pkg{AtlasMapper}: mapping between coordinates and anatomical labels for a \emph{single} atlas.
    \item \pkg{MultiAtlasMapper}: coordinated mapping across \emph{multiple} atlases, returning per-atlas labels and facilitating cross-atlas comparison.
    \item \pkg{BatchAtlasMapper}: high-throughput processing of \emph{sets} of coordinates (e.g., many participants or conditions), wrapping the previous mappers with batching, caching, and progress reporting.
\end{itemize}
All three support single points or large batches; convert bidirectionally between coordinates, voxel indices, and region labels; and automatically infer hemisphere from region labels. \emph{Example: 500 lesion coordinates can be batch-processed across AAL and Destrieux in one query, with outputs labeled consistently.}

Edge cases are resolved via nearest-neighbor interpolation so coordinates that do not overlap labeled voxels map to the nearest anatomical region. For volume atlases, we precompute the set of labeled voxels $V \subset \mathbb{Z}^3$ and build a KD-tree over their MNI coordinates. Given an input location $x$, if $x$ is unlabeled we return the label at
\[
v^\ast = \arg\min_{v \in V} \bigl\| x - \operatorname{mni}(v) \bigr\|_2
\]
and report the distance
\[
d = \bigl\| x - \operatorname{mni}(v^\ast) \bigr\|_2.
\]
For surface atlases, we index labeled vertices and build a KD-tree over their 3D positions; for vertex-to-triangle ambiguities, we prefer the smallest geodesic distance when available, otherwise the Euclidean distance. KD-trees, centroids, and label maps are cached in a user-configurable directory and serialized for reuse (batch API/CLI), enabling large-scale, cross-atlas mapping on standard hardware.

\subsection{Datasets (Neurosynth \& NeuroQuery) and Study Search}
Through \pkg{fetch\_datasets}, \pkg{deduplicate\_datasets}, and \pkg{search\_studies}, Coord2Region provides a NiMARE-backed integration with large-scale meta-analytic resources (Neurosynth, NeuroQuery). This module links atlas-derived coordinates or region labels to relevant published studies and associated cognitive terms. \emph{Example: a researcher entering coordinates in the posterior cingulate cortex receives a curated list of studies relating this region to memory and default mode network activity, with DOIs and abstracts included when available.}

\noindent
Missing bibliographic fields (e.g., DOI, journal, abstract) are automatically backfilled from PubMed and CrossRef. The resulting standardized study records include identifiers (PMID and/or DOI when available), titles, abstracts, and the source dataset, spanning more than 14{,}000 articles via Neurosynth and 13{,}000 via NeuroQuery. Cross-source deduplication ensures that overlapping entries are merged into a unique study set-for instance, if both datasets return the same hippocampal study, it is listed only once, reducing redundancy in literature searches.

\noindent
By tying anatomical results directly to the literature, this module enables rapid functional interpretation and efficient study discovery. All outputs can be exported to JSON or CSV for downstream meta-analysis, integration into custom pipelines, or sharing with collaborators.

\subsection{LLM utilities}
\pkg{AIModelInterface} provides a unified wrapper for multiple LLM providers (OpenAI, Anthropic, Google Gemini, OpenRouter, Hugging Face). It implements robust retries with backoff and timeouts, asynchronous batch generation, token-aware streaming, an extensible library of prompt templates (with support for user-defined templates), and LRU caching to avoid redundant calls. For figures, it can compose MNI-space overlays with AI-generated region sketches and apply watermarking.

\noindent
The API accepts single or batched inputs (coordinates or region labels) and returns structured outputs (e.g., JSON, CSV) and LaTeX-ready text. This module standardizes narrative reporting (summaries, captions, methods blurbs) across studies, reduces manual literature-synthesis effort, and scales to large datasets. Caching and templating improve reproducibility and consistency across runs and providers, while asynchronous and streaming modes shorten end-to-end pipelines and simplify integration in notebooks, scripts, and the CLI.

\subsection{Pipeline and CLI commands}
Coord2Region supports multiple usage levels to accommodate users with different technical backgrounds. At the lowest level, users may directly call individual functions (e.g., \texttt{mni\_to\_region\_name}, \texttt{get\_studies\_for\_coordinates}, \texttt{generate\_summary}) to build custom workflows. \emph{Example: a methods developer can script a pipeline that maps 300 MNI coordinates to AAL atlas regions, retrieves associated literature, and saves the results as a CSV for meta-analysis.}

\noindent
For higher-level use, the \texttt{Pipeline} API (in \texttt{pipeline.py}) chains the key processing steps—coordinate mapping, dataset fetching and deduplication, study retrieval, image generation, and summarization—into a single coherent workflow that returns consistent outputs (region names, study lists, images, summaries). Alternatively, the command-line interface (\texttt{CLI} in \texttt{cli.py}) exposes the same functionality via commands such as \texttt{coords-to-summary}, \texttt{coords-to-image}, \texttt{coords-to-insights}, \texttt{region-to-insights} and \texttt{coords-to-study}, and can optionally consume YAML configuration files to facilitate reproducible execution and sharing of workflows.
\begin{figure}[h!]
  \centering
  \includegraphics[width=0.92\linewidth]{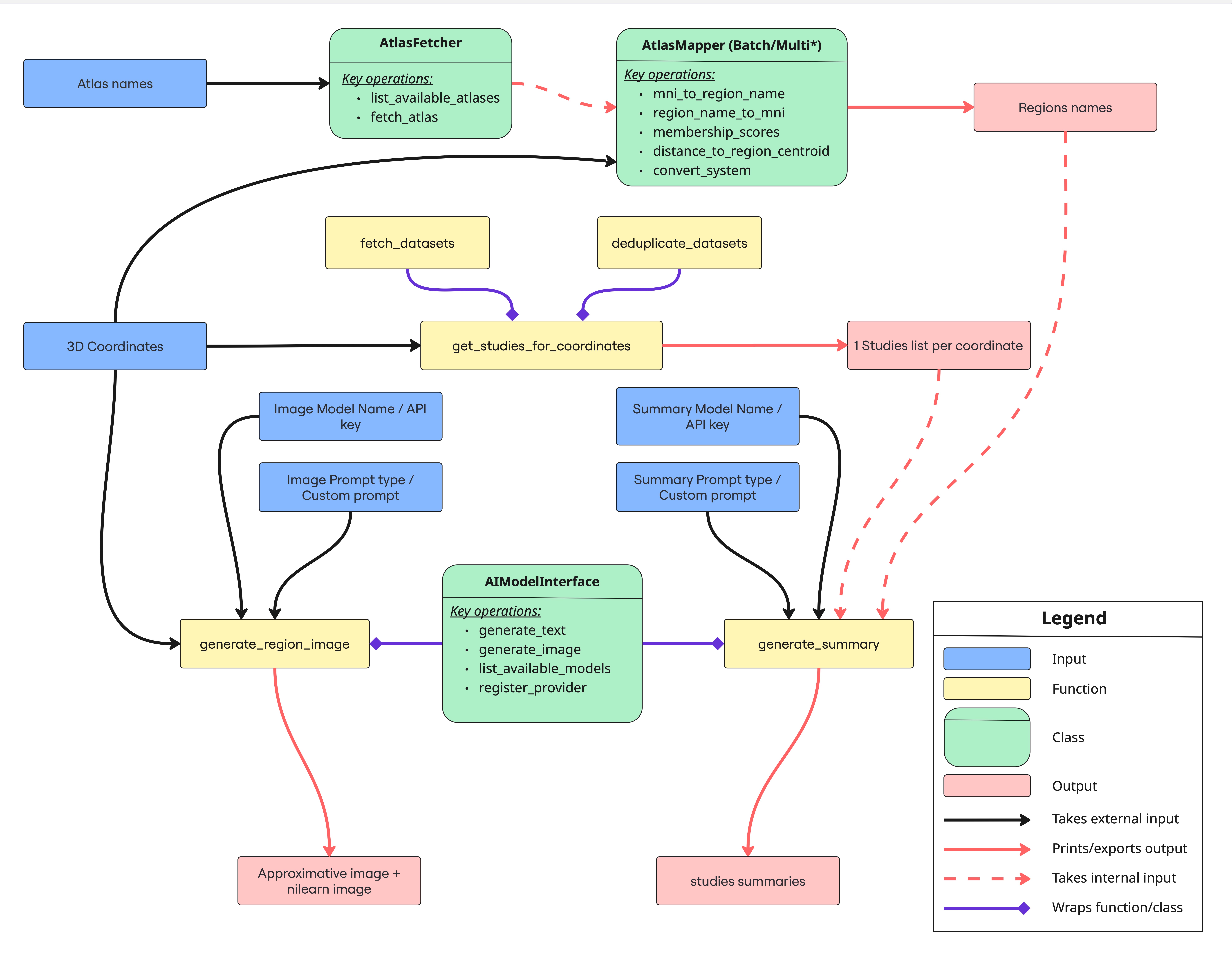}
  \caption{\textbf{Overview of the Coord2Region workflow}. The pipeline starts with user-provided inputs (blue), such as atlas names, 3D coordinates, or model specifications. Core functions (yellow) orchestrate data retrieval, mapping, and analysis, relying on dedicated classes (green) like \texttt{AtlasFetcher}, \texttt{AtlasMapper} (Batch/Multi*), and \texttt{AIModelInterface}. Internal data flow and dependencies are shown with dashed red arrows, while external inputs and outputs follow solid black and red connections. The system produces outputs (pink), including mapped region names, study summaries, and approximate images. In the case where region names are provided as input, the workflow is similar but restricted to a single atlas. The “Batch/Multi*” label indicates the existence of \texttt{BatchAtlasMapper} (handling multiple coordinates simultaneously) and \texttt{MultiAtlasMapper} (using multiple atlases at once); both build on \texttt{AtlasMapper} but are omitted for simplicity.}
  \label{fig:pipeline}
\end{figure}

\noindent
Users can therefore choose to work purely in Python code, via the high-level pipeline API, or through the CLI (with or without YAML), and can transition between modes as their needs evolve. This tiered design ensures accessibility for a wide range of researchers, from developers building advanced pipelines to clinical scientists seeking quick coordinate-to-region mappings.

\noindent
Figure~\ref{fig:pipeline} illustrates the architecture and data flow of the Coord2Region pipeline. The process begins when the user supplies either atlas names or 3D coordinates as input. The \texttt{AtlasFetcher} component first retrieves or lists available atlases. Next, \texttt{AtlasMapper} (or its \texttt{BatchAtlasMapper} / \texttt{MultiAtlasMapper} variants) converts between MNI coordinates and region names, computes distances to centroids or labeled voxels, and performs system conversions as needed. Once mapping is complete, the system loads cached data or fetches datasets, deduplicates overlapping entries, and \texttt{get\_studies\_for\_coordinates} retrieves a list of relevant studies for each coordinate.

\noindent
In parallel, the user’s selection of image or summary models and prompt templates is passed to \texttt{AIModelInterface}, which wraps methods such as \texttt{generate\_text} and \texttt{generate\_image}. The pipeline invokes \texttt{generate\_region\_image} to produce approximate visual representations of mapped regions and \texttt{generate\_summary} to synthesize textual summaries based on the retrieved study lists. The final outputs include region names, study lists per coordinate, approximate region images, and study summaries. When region names are provided instead of coordinates, the workflow is analogous, with the restriction that a single atlas is used.

\noindent
The label “Batch/Multi*” in the \texttt{AtlasMapper} block denotes that \texttt{BatchAtlasMapper} (for handling multiple coordinates) and \texttt{MultiAtlasMapper} (for handling multiple atlases) are supported; both extend or wrap the base \texttt{AtlasMapper} and are omitted from the diagram for clarity. Additional classes, utilities, error-handling, and caching modules exist in the full implementation but are not shown.

\section{Web Interface: Config Builder and Cloud Runner}

\begin{figure}[htbp]
  \centering
  \includegraphics[width=0.8\textwidth]{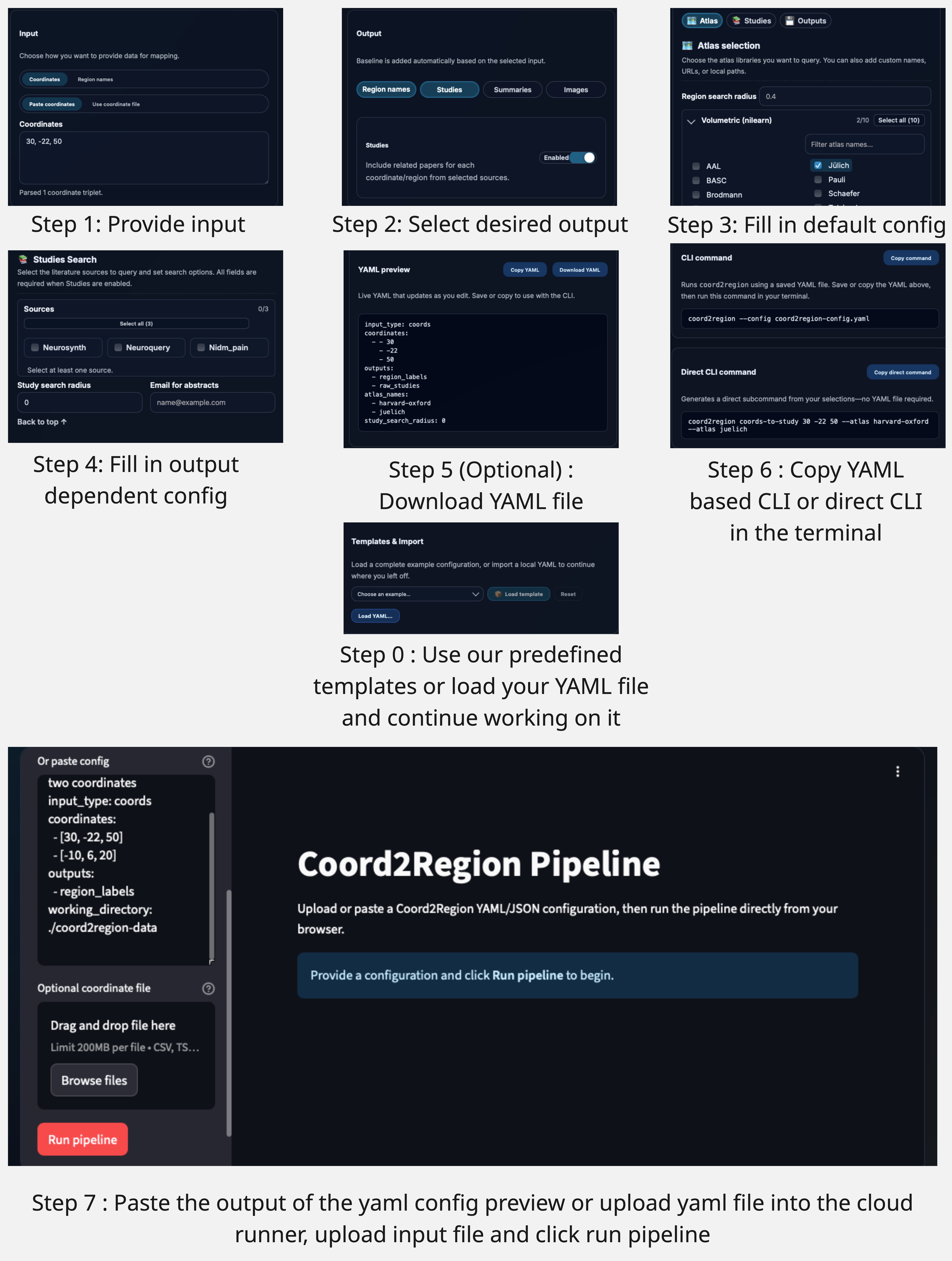}
  \caption{Coord2Region web interface workflow. Users begin by providing input (e.g., coordinates or region names) in Step~1, then choose the desired output type (Step~2). In Step~3 they specify atlas and output settings, and in Step~4 configure output-related parameters. The live YAML preview updates as settings change (Step~5), after which the user can either copy a CLI command or download a YAML configuration (Step~6). Finally (Step~7), the user pastes or uploads the YAML into the Cloud Runner interface, supplies any remaining inputs, and runs the pipeline in the browser.}
  \label{fig:web_interface_workflow}
\end{figure}

Coord2Region also offers a fully functional \textbf{web interface}\footnote{\url{https://babasanfour.github.io/Coord2Region/}}(Figure~\ref{fig:web_interface_workflow}) that lowers the barrier to entry for non-programmers and supports reproducible workflows directly in the browser. At the heart of this interface is the \emph{Config Builder}, which is powered by JSON Schema–driven forms and presents users with dynamic fields for specifying input mode (coordinates or region names), atlas selection, data sources, output options (studies, summaries, images), and model or prompt parameters.

\noindent
The Config Builder guides users through the following steps:
\begin{enumerate}
  \item Enter coordinates (single or batch) or region names.
  \item Select one or more atlases (e.g., AAL, Harvard--Oxford, Destrieux).
  \item Choose desired outputs such as anatomical labels, linked studies, summaries, or images.
  \item Configure model and output options (e.g., number of studies retrieved, whether to generate summaries, output formats).
\end{enumerate}

\noindent
As users complete the form, live validation ensures that only configurations consistent with the underlying schema are allowed. The interface also displays a live YAML preview and a corresponding CLI command snippet, enabling users to download the YAML, copy the command, or directly invoke CLI execution based on their selections.

\noindent
Complementing the Config Builder is the \emph{Cloud Runner}, a Streamlit-based web app (e.g., deployed via Hugging Face Spaces) that wraps the full Coord2Region pipeline. Users can submit inputs, choose atlases and options, and run mapping, study retrieval, summarization, and image generation entirely in the browser, without requiring a local installation. Because the Cloud Runner uses the same core logic as the CLI and pipeline, results remain consistent across interfaces. The current Web Runner, hosted on Hugging Face Spaces, is intended as a demonstration environment and supports single-coordinate or small-scale queries; high-throughput runs should use the local CLI and Python interfaces.

\noindent
Together, the Config Builder and Cloud Runner enable a seamless, end-to-end web experience: users can interactively build valid YAML configurations, execute analyses in the cloud, and, if desired, transition to CLI or Python usage using the same configuration files. The schema-driven design ensures that any future additions to configuration options are automatically reflected in the web UI, keeping all modes in sync with minimal maintenance.

\section{Discussion}
Coord2Region addresses a central bottleneck in coordinate-based neuroimaging: translating 3D coordinates into consistent anatomical labels with direct links to supporting literature. Widely used visualization tools (e.g., MRIcroGL, NiiVue) and meta-analytic platforms (e.g., Neurosynth, NeuroQuery, NiMARE) provide powerful capabilities, but none offer a unified workflow that integrates atlas-based labeling, study retrieval, and automated summaries.

\noindent
Compared with manual coordinate lookups in GUI software, Coord2Region offers:
\begin{itemize}
  \item \textbf{Speed and scalability.} Large datasets of hundreds or thousands of coordinates can be batch-processed within minutes using KD-tree acceleration and caching.
  \item \textbf{Consistency.} Multi-atlas queries ensure that results are reproducible and comparable across studies, with explicit reporting of nearest-region distances for ambiguous cases.
  \item \textbf{Literature integration.} Direct links to peer-reviewed publications reduce the time spent manually searching PubMed or Google Scholar for functional interpretation.
\end{itemize}

\noindent
Compared with meta-analytic frameworks alone, Coord2Region fills the “glue layer” by providing coordinate-level entry points. A researcher can start with a single activation peak or lesion coordinate and quickly move to literature-backed summaries without needing full statistical maps.

\noindent
The inclusion of LLM-powered summaries and illustrative images extends this pipeline by offering optional interpretive aids. These outputs are not intended to replace systematic review, but instead provide a useful first-pass synthesis, particularly in exploratory or clinical contexts. \emph{For example, a clinician localizing electrode sites may use the AI-generated summary to obtain a quick overview of cognitive functions associated with a mapped region before examining the primary literature in detail.}

\noindent
Overall, Coord2Region complements existing visualization and meta-analytic ecosystems by serving as an automation and reproducibility layer. It lowers the barrier to consistent coordinate labeling, accelerates functional interpretation, and reduces human error, while maintaining compatibility with established neuroimaging libraries (nibabel, nilearn, MNE-Python) and databases (Neurosynth, NeuroQuery, NiMARE).

\noindent
Beyond its Python API, Coord2Region is deliberately designed to be accessible to users with little or no programming experience. The web-based Config Builder and Cloud Runner provide a point-and-click interface for defining, running, and sharing analyses. By exposing all key options—input mode, atlas selection, output types, and model parameters—through JSON Schema–driven forms, the web interface allows users to construct valid configurations without writing code or editing YAML manually. Live validation and immediate YAML/CLI previews further reduce errors and support transparent reporting.

\noindent
Similarly, the command-line interface offers a low-friction entry point for non-programmers who are comfortable running simple terminal commands. High-level commands (e.g., \texttt{coords-to-summary}, \texttt{coords-to-study}, \texttt{coords-to-insights}) encapsulate the full pipeline behind a small set of arguments, and YAML configuration files can be reused across machines, projects, and collaborators. Together, the web interface and CLI make it possible for students, clinicians, and other domain experts to benefit from automated atlas labeling and literature linkage without needing to develop custom Python scripts, while still remaining fully compatible with more advanced, code-centric workflows.

\noindent
Future work will extend the package with probabilistic atlases, M/EEG sensor support, and automated integration of new literature as it becomes available. By situating coordinate mapping within a reproducible, scriptable, and literature-linked workflow, Coord2Region supports more reliable and transparent reporting in neuroimaging research.

\section{Reproducibility \& Availability}

\noindent\textbf{Open source and licensing.}
\pkg{Coord2Region} is open source and developed publicly on GitHub. Issues, discussions, examples, and continuous-integration (CI) logs are available to support transparent development and community review. The project welcomes external contributions via standard pull-request workflows, with coding style and contribution guidelines documented in the repository.

\noindent\textbf{Core dependencies.}
The package targets Python~$\geq$~3.10 and relies on widely adopted scientific libraries:
\texttt{numpy}, \texttt{scipy}, \texttt{pandas}, \texttt{nibabel}, \texttt{nilearn}, and (for surface workflows) \texttt{MNE-Python}. The command-line interface is implemented using \texttt{argparse} and standard library modules.

\noindent\textbf{Optional integrations.}
For literature retrieval and decoding, \pkg{Coord2Region} interoperates with \texttt{NiMARE} and its supported resources (e.g., Neurosynth, NeuroQuery). LLM-powered summaries and image synthesis are optional features enabled via provider-specific Python packages (e.g., OpenAI, Anthropic, Google Gemini, OpenRouter SDKs, or local Hugging Face backends). These integrations are isolated behind the \texttt{AIModelInterface}, so that core atlas mapping remains provider-agnostic and usable without any AI keys.

\noindent\textbf{Determinism and caching.}
Atlas volumes, surface annotations, centroids, and KD-trees are cached under a user-configurable data directory to accelerate repeated runs. Package and atlas versions are recorded in outputs, and random seeds are fixed where applicable. All pipeline steps can export machine-readable artifacts (JSON/CSV) alongside human-readable reports (PDF) to support reuse, inspection, and audit.

\noindent\textbf{Reproducible interfaces.}
Every capability available in Python is mirrored in the \texttt{coord2region} CLI (e.g., \texttt{coords-to-atlas}, \texttt{coords-to-summary}, \texttt{coords-to-insights}, \texttt{region-to-coords}), enabling both scripted and interactive use. Examples demonstrate end-to-end calls, expected inputs/outputs, and how to pin software and data versions for long-term reproducibility.

\noindent\textbf{Data and third-party resources.}
When fetching atlases or meta-analytic datasets, the package preserves original licenses and citation information. Users are encouraged to cite the relevant atlas and dataset sources when publishing results. Installation instructions, minimal examples, and CI status badges are linked from the README and the online documentation.

\section{Limitations}
Atlas choice and resolution inherently influence derived labels; surface/volume mismatches can introduce edge effects and partial-volume artifacts. Nearest-region fallbacks introduce distance-dependent uncertainty, so Coord2Region reports distances and allows user-defined thresholds to flag or exclude low-confidence labels. LLM-based summaries may reflect provider biases, incomplete coverage of the literature, and model drift over time; to mitigate this, we record provenance (model name, version, prompts) and support offline or local models when available. Users should avoid over-interpreting labels as precise neuroanatomical ground truth, should treat AI-generated text and images as heuristic aids rather than definitive evidence, and must respect dataset licenses, citation requirements, and human-subjects constraints when using or sharing results.

\section{Future Directions}
Coord2Region actively encourages community participation and future enhancements. Current priorities for ongoing development include:
\begin{itemize}
  \item \textbf{Richer visualization.} Integrating more interactive views of coordinate mappings onto brain surfaces and volumes (e.g., web-based 3D viewers, hoverable annotations, montage exports).
  \item \textbf{Expanded atlas and modality support.} Adding probabilistic and functional atlases, improved handling of subject-specific spaces, and first-class support for M/EEG sensor and source-space coordinates.
  \item \textbf{Enhanced AI workflows.} Incorporating more robust prompting strategies, support for retrieval-augmented LLM summaries grounded in the fetched literature, and tools for systematically comparing outputs across models.
  \item \textbf{Automated literature updates.} Streamlining periodic refresh of meta-analytic datasets and bibliographic metadata so that coordinate-to-literature mappings remain current as new studies are published.
\end{itemize}
The project welcomes contributions in the form of new atlases, additional example notebooks, improvements to documentation, bug reports, feature requests, and pull requests implementing new functionality. Community feedback will guide the prioritization of future features and integrations.

\section{Documentation and Package Repository}
For further information, detailed guidelines, and usage examples, users can consult the online documentation at
\url{https://coord2region.readthedocs.io/en/latest/} and explore the package repository at
\url{https://github.com/BabaSanfour/Coord2Region} for source code, issue tracking, and collaborative development. These resources are continuously updated to support and enhance the user experience, including tutorials, API reference, and release notes.

\section{Acknowledgments}
We gratefully acknowledge the foundational contributions of the developers of nilearn, nibabel, and MNE-Python, whose open-source neuroimaging tools made Coord2Region possible. We also thank the NiMARE community for maintaining and expanding critical neuroimaging meta-analytic resources that greatly enhance Coord2Region’s capacity for functional interpretation. The authors would additionally like to thank AntiCafe Loft for their support.

\bibliographystyle{plainnat}
\bibliography{refs}
\end{document}